\documentclass[aps,nofootinbib,superscriptaddress, showpacs,preprintnumbers,twocolumn]{revtex4-1}

\usepackage{graphicx}
\usepackage{dcolumn}
\usepackage{bm}
\usepackage{eurosym}
\usepackage{amsmath}
\usepackage{bm}
\usepackage{amsfonts}
\usepackage{amssymb}
\usepackage{hyperref}
\usepackage{mathtools}
\usepackage{color}
\usepackage{dcolumn}
\usepackage{bm}
\usepackage{caption}
\usepackage{subcaption}
\usepackage{epstopdf}
\usepackage{xcolor}

\begin{document}

\title{The Exact Relativistic Scalar Quasibound States of The Dyonic Kerr-Sen Black Hole: Quantized Energy, and Hawking Radiation}

\author{David Senjaya}
\email{davidsenjaya@protonmail.com}

\author{Piyabut Burikham}%
 \email{piyabut@gmail.com}
 \affiliation{High Energy Physics Theory Group, Department of Physics,
Faculty of Science, Chulalongkorn University, 254 Phayathai Road, Pathumwan, 10330, Bangkok, Thailand}
\author{Tiberiu Harko}
\email{tiberiu.harko@aira.astro.ro}
\affiliation{Department of Physics, Babes-Bolyai University,1 Mihail Kogalniceanu Street,400084, Cluj-Napoca,Romania}
\affiliation{Astronomical Observatory,19 Ciresilor Street,400487,Cluj-Napoca,Romania}

\date{\today}

\begin{abstract}
We consider Klein-Gordon equation in the Dyonic Kerr-Sen black hole background, which is the charged rotating axially symmetric solution of the Einstein-Maxwell-Dilaton-Axion theory of gravity. The black hole incorporates electric, magnetic, dilatonic and axionic charges and is constructed in 3+1 dimensional spacetime. We begin our investigations with the construction of the scalar field's governing equation, i.e., the covariant Klein-Gordon equation. With the help of the ansatz of separation of variables, we successfully separate the polar part, and find the exact solution in terms of Spheroidal Harmonics, while the radial exact solution is obtained in terms of the Confluent Heun function. The quantization of the quasibound state is done by applying the polynomial condition of the Confluent Heun function that gives rise to discrete complex-valued energy levels for massive scalar fields. The real part is the scalar field relativistic quantized energy, while the imaginary part represents the quasibound states's decay. We present all of the sixteen possible exact energy solutions for both massive and massless scalars. We also present the investigation the Hawking radiation of the Dyonic Kerr-Sen black hole's apparent horizon, via the Sigurd-Sannan method by making use of the obtained exact scalar wave functions. The radiation distribution function, and the Hawking temperature are also obtained.
\end{abstract}

\keywords{EMDA super gravity, Kerr-Sen black hole, Klein-Gordon equation, quasibound states, Hawking radiation}

\maketitle





\section{Introduction}
Einstein's theory of general relativity is a successful successor to the old Newtonian theory of gravity. It describes the gravitational field as a curvature effect in
spacetime, determined by the mass of a body. The motion of particles in the curved spacetime is governed by the geodesic equations, which are directly related to the spacetime metric. Some of the gravitational effects predicted by general relativity, such as the perihelion precession of Mercury, the bending of light, the gravitational redshift of the distant stars, the gravitational waves, and the existence of black holes, have been confirmed observationally, or experimentally \cite{Hobson,Will,Abbott,Abbott1,EHTC}.

However, despite its remarkable phenomenological success, Einstein's theory has also several problems, such as, for example,  the existence of singularities in its black hole and cosmological solutions. Also, dark matter and dark energy are required to be added by hand inside the source term ($T_{\mu\nu}$), in order to explain the flat galaxy rotation curves, and the accelerated expansion of the Universe. The need of modifying the Einstein's general relativity then gave birth to various extended theories of gravity, such as the Kaluza-Klein theory, $f(R), f(T)$, $f\left(R,L_m\right)$, $f(R,T)$ modified gravities, the Gauss-Bonnet theory, the Scalar-tensor-vector gravity,  Lovelock theory,  etc \cite{Buchdahl, Linder, H1,H2,  Moffat2006,Moffat2007,Clifton,papantonopoulos, Petrov}.

In this work, we will focus on the string inspired Einstein-Maxwell-dilaton-axion (EMDA) theory of gravity, which belongs to the scalar-vector-tensor supergravity family. The EMDA theory generalizes the Einstein-Maxwell theory by introducing a coupling between the Maxwell electromagnetic field tensor $F_{\mu\nu}$,  the scalar dilaton field $\xi$ and the pseudo scalar axion field $\phi$, and it can be constructed from the following four dimensional effective action \cite{Sen,Baner}
\begin{multline}
S_{EMDA}=\frac{1}{16\pi}\int\left[R-2\partial_\mu\xi\partial^\mu\xi\right. \\ \left.-\frac{1}{3}H_{\rho\sigma\delta}H^{\rho\sigma\delta}+e^{-2\xi}F_{\alpha\beta}F^{\alpha\beta}\right]\sqrt{-g}d^4x,
\end{multline}
where $R$ is the Ricci scalar and $g$ is the metric tensor determinant. The Maxwell electromagnetic field tensor $F_{\mu\nu}$ is defined as the partial derivatives of the $U(1)$ gauge field $A_\mu$ as follows,
\begin{equation}
F_{\mu\nu}=\partial_\mu A_\nu-\partial_\nu A_\mu.
\end{equation}
Moreover, $H^{\rho\sigma\delta}$ is the Kalb-Ramond field tensor that is written in terms of the pseudo-scalar axion field $\phi$ according to,
\begin{equation}
H_{\alpha\beta\delta}=\frac{1}{2}e^{4\xi}\varepsilon_{\alpha\beta\delta\gamma}\partial^\gamma \phi.
\end{equation}

Recently, Wu et al. \cite{Wu} did find a novel exact rotating black hole solution of the EMDA gravity that depends on four charges, the i.e. electric, magnetic, dilaton and axion charges, called the Dyonic Kerr-Sen black hole. The Dyonic Kerr-Sen black hole is a generalization of the non-dyonic Kerr-Sen black hole, found in \cite{Sen}. The presence of the additional magnetic, dilaton and axion fields in the theory also generalizes the electro-vacuum rotating black hole solution of the Einstein-Maxwell theory of gravity, the Kerr-Newman black hole solution \cite{Kerr}. Due to its generality, in this work, the aforementioned Dyonic Kerr-Sen solution becomes our topic of interest in order to investigate the scalar quasibound states and the Hawking radiation, respectively.

The behavior of the relativistic scalar fields in curved spacetime is an important subject of research, which may give some insights into the quantum theory of gravity. In the presence of matter-charge distribution, the surrounding spacetime will be curved, and the scalar field-gravity interaction is then investigated by looking at the scalar field behavior in the curved spacetime backgrounds. The behavior of the relativistic scalar fields in curved spacetimes is mathematically described by the covariant Klein-Gordon equation, in which the gravitational field is introduced via the metric tensor $g_{\mu\nu}$. The investigation of the Klein-Gordon equation in the black hole background may open up a new understanding of the quantum theory of gravity. Instead of trying to consider the quantization of the general
relativity, in the present work we take the opposite approach, by "gravitizing" the quantum mechanics, to formulate it in line with the principles of the theory of general relativity.

After the gravitational wave signal of a binary black hole merger was finally detected for the first time on 14 September 2015 \cite{205}, quite recently, the Hawking radiation of an optical black hole analog \cite{drori} was also finally observed. This makes the investigation of the black hole spectroscopy a new emerging field of major physical interest. The quasibound states~(QBS), quasinormal modes~(QNMs), and shadows of black holes are among the most interesting characteristics of such astrophysical objects, and they appear in the observationally measurable spectra that are released as the particles cross into the black hole \cite{100}.

The quasibound states are relativistic quasistationary resonances outside the black hole's event horizon, localized in the black hole's finite gravitational potential well. In contrast to the electron's effective potential in the hydrogen atoms, scalar quasibound states are leaking and crossing into the black hole, causing the spectrum to be complex valued, similar to the case of damped oscillations, where the real part is the scalar field's quantized energy, and the imaginary part determines the stability of the system. It is possible, in principle, to extract the information about the central black hole, as well as to validate some modified theories of gravity, by measuring the frequency of the quasibound states \cite{100}. Analogously to atomic transitions emitting photons,  transitions of axions around black holes between different energy levels emit gravitons \cite{32a}.

However, due to the complexity of the equations involved, especially the radial equation, analytical methods were used less often, and only for some very special problems \cite{Anal1, Anal2, Anal3}. The vast majority of these studies made use of numerical techniques such as asymptotical analysis, WKB, and continued fraction to investigate the specific task at hand. However, very recently, \cite{Vier22,senjaya1,senjaya2,senjaya3,senjaya4,senjaya5,senjaya6,senjaya7} have successfully worked out, and presented some novel exact scalar quasibound states solutions in various black hole backgrounds, and obtained the radial exact solutions of the Klein-Gordon equation in terms of the Confluent Heun and the General Heun functions. The polynomial condition of the Heun functions leads to the quasibound states' energy quantization.

In this work, we present in detail the analytical derivations of both massive and massless scalar quasibound states of the  exact solutions in the Dyonic Kerr-Sen black hole background. The exact wave function is found in terms of the harmonic function as the temporal solution, the Spheroidal harmonics as the angular solution, and the Confluent Heun functions as the radial solutions. The expression of the relativistic quantized energy levels is obtained after applying the polynomial condition of the Confluent Heun function. We also investigate the wave function of the quasibound states in the two extreme regions, i.e., close to the black hole horizon, and at infinity, respectively,  and find out that they behave like an ingoing wave, close to the black hole horizon, and vanishing far away from the horizon. In the last Section, using exact radial solutions, the Hawking radiation of the black hole's apparent horizon is investigated, and the Hawking temperature is obtained.

\section{The Dyonic Kerr-Sen Metric}

In the present Section we briefly review the basic properties of the Dyonic Kerr-Sen metric, we obtain the singular points of the line element, and we calculate the inverse of the metric tensor coefficients.

\subsection{The Metric}

The line element of rotating black hole's exterior can be written in either the Cartesian (Kerr-Schild) coordinates, the Boyer-Lindquist (Schwarzschild-like) coordinates, or the Eddington-Finkelstein (Tortoise) coordinates. In this work, the Dyonic Kerr-Sen spacetime in the Boyer-Lindquist coordinate system is adopted. The line element is given as follows \cite{Wu,Jana},
\begin{multline}
ds^2=-\left[1-\frac{r_s\left(r-d\right)-r^2_D}{\rho^2}\right]c^2dt^2\\
-2\frac{r_s\left(r-d\right)-r^2_D}{\rho^2}a{{\sin }^{{\rm 2}} \theta\ }d\phi cdt+\frac{\rho^2}{\Delta }dr^2+\rho^2d\theta^2\\+\left[r\left(r-2d\right)-k^2+a^2+\frac{r_s\left(r-d\right)-r^2_D}{\rho^2}a^2{{\sin }^{{\rm 2}} \theta\ }\right]\times \\{{\sin }^{{\rm 2}} \theta\ } d\phi^2, \label{metric}
\end{multline}
where,
\begin{gather}
\rho^2=r\left(r-2d\right)-k^2+a^2{{\cos }^{{\rm 2}} \theta\ },\\
\Delta =r\left(r-2d\right)-r_s\left(r-d\right)-k^2+a^2+r^2_D.
\end{gather}
We have also introduced the notations,
\begin{gather}
r^2_D=\frac{G\left(Q^2_E+Q^2_M\right)}{4\pi \epsilon_0 c^4}=Q^2+P^2,\\
k=2\frac{PQ}{r_s},\\
d=\frac{P^2-Q^2}{r_s},\\
d^2+k^2={\left(\frac{P^2+Q^2}{r_s}\right)}^2\leftrightarrow r_s\sqrt{d^2+k^2}=r^2_D,\\
a=\frac{J}{M},
\end{gather}
where $M, J, Q, P, d, k$ are respectively the black hole's mass, spin, electric, magnetic/dyonic, dilaton and axion charge.

The Boyer-Lindquist coordinates are connected with the Cartesian coordinates by the following transformation,
\begin{gather}
x=\sqrt{r\left(r-2d\right)+a^2+r^2_D-k^2+dr_s}{\sin  \theta{\cos  \phi\ }\ },\\
y=\sqrt{r\left(r-2d\right)+a^2+r^2_D-k^2+dr_s}{\sin  \theta{\sin  \phi\ }\ },\\
z=\sqrt{r\left(r-2d\right)}\;{\cos  \theta\ }.
\end{gather}

The Dyonic Kerr-Sen geometry is singular when $\rho^2=0$ and $\Delta {\rm =0}$. The first condition,
\begin{equation}
\rho^2=0=r\left(r-2d\right)-k^2+a^2{{\cos }^{{\rm 2}} \theta\ },
\end{equation}
is satisfied at $\theta=\frac{\pi}{2}$. By using the Cartesian-Boyer-Lindquist coordinates relationship,
\begin{equation}
 x^2+y^2=r\left(r-2d\right)+a^2+r^2_D-k^2+r_sd,
\end{equation}
we obtain,
\begin{equation}
r\left(r-2d\right)-k^2=0\leftrightarrow x^2+y^2=a^2+r^2_D+r_sd.
\end{equation}

Here we obtain a ring singularity on the Cartesian $x-y$ plane, with radius $\sqrt{a^2+r^2_D+r_sd}$, measured from the coordinate's origin. The second singularity, $\Delta {\rm =0,}$ is related to the one way surfaces, i.e., the black hole's horizons, determined by the condition $g_{rr}=\infty $. The positions of the horizons can be obtained by solving the following quadratic equation,
\begin{gather}
 \Delta {\rm =0=}r\left(r-2d\right)-r_s\left(r-d\right)-k^2+a^2+r^2_D,\\
 r_{\pm }=\frac{r_s}{2}+d\pm \sqrt{{\left(\frac{r_s}{2}\right)}^2+d^2+k^2-\left(a^2+r^2_D\right)}, \label{rpm}
\end{gather}
where $r_+$ and $r_-$ are  the outer and inner horizons, respectively.

It is important to mention that by taking $d=0$ in the expression of $k$, we reobtain the non-dilatonic, axionic dyonic black hole, and the non-axionic, dilatonic dyonic black hole solutions. By  taking $P$ equal to zero, we obtain the Kerr-Newman black hole, which is the most general electro-vacuum black hole solution of the Einstein-Maxwell theory. By taking $P,a$ equal to zero, we reobtain the static spherically symmetric Reissner-Nordstr\"om black hole. By taking as zero all of the Dyonic Kerr-Sen black hole's parameters, except the mass, we obtain the basic Schwazschild black hole.

In order to find  the inverse of the metric of the Dyonic Kerr-Sen spacetime, as a first step we express the line element \eqref{metric} as follows,
\begin{gather}
g_{\mu\nu}=\left( \begin{array}{cccc}
-\left[1-\frac{r_s\left(r-d\right)-r^2_D}{\rho^2}\right] & 0 & 0 & g_{c\phi}  \\
0 & \frac{\rho^2}{\Delta } & 0 & 0 \\
0 & 0 & \rho^2 & 0 \\
g_{c\phi} & 0 & 0 & g_{\phi\phi} \end{array}
\right),
\end{gather}
where,
\begin{equation}
g_{c\phi}= -\frac{r_s\left(r-d\right)-r^2_D}{\rho^2}a{\sin }^{{\rm 2}} \theta,
\end{equation}\vspace{-.5cm}
\begin{multline}
g_{\phi\phi}=\Bigg[r\left(r-2d\right)-k^2+a^2 \\ \left.+\frac{r_s\left(r-d\right)-r^2_D}{\rho^2}a^2{{\sin }^{{\rm 2}} \theta\ }\right]{{\sin }^{{\rm 2}} \theta\ }.
\end{multline}

Since we need to calculate the determinant of the metric tensor, let us first modify the metric tensor to be a block matrix as follows,
\begin{align}
g_{\mu\nu}&=\left( \begin{array}{cccc}
\frac{\rho^2}{\Delta } & 0 & 0 & 0 \\
0 & \rho^2 & 0 & 0 \\
0 & 0 & -\left[1-\frac{r_s\left(r-d\right)-r^2_D}{\rho^2}\right] & g_{c\phi} \\
0 & 0 & g_{c\phi} & g_{\phi\phi} \end{array}
\right)\\&=\left( \begin{array}{cc}
\tilde{A} & 0 \\
0 & \tilde{B} \end{array}
\right).
\end{align}

The determinant of the metric tensor can then be calculated as follows,
\begin{equation}
{\det  \left(g_{\mu\nu}\right)\ }=g=\det  \left(\tilde{A}\right)\det  \left(\tilde{B}\right),
\end{equation}
and after some lines of straightforward algebra, we obtain,
\begin{equation}
g=-\rho^4{{\sin }^{{\rm 2}} \theta\ }.
\end{equation}

\subsection{The Metric Inverse}

Now we will proceed to calculate the inverse of the metric tensor. Using the metric in block matrix form, the inverse is calculated block per block as follows,
\begin{gather}
g_{\mu\nu}=\left( \begin{array}{cc}
\tilde{A} & 0 \\
0 & \tilde{B} \end{array}
\right)\to g^{\mu\nu}=\left( \begin{array}{cc}
{\tilde{A}}^{-1} & 0 \\
0 & {\tilde{B}}^{-1} \end{array}
\right),\\
{\tilde{A}}^{-1}=\left( \begin{array}{cc}
\frac{\Delta }{\rho^2} & 0 \\
0 & \frac{1}{\rho^2} \end{array}
\right),\\
{\tilde{B}}^{-1}=\frac{1}{\left|\tilde{B}\right|}\left( \begin{array}{cc}
g_{\phi\phi} & -g_{c\phi} \\
-g_{c\phi} & -\left[1-\frac{r_s\left(r-d\right)-r^2_D}{\rho^2}\right] \end{array}
\right),\\
\left|\tilde{B}\right|=\frac{-g}{\left(\frac{\rho^4}{\Delta }\right)}=-\Delta {{\sin }^2 \theta\ }.
\end{gather}

After finding the inverse of each block, the metric tensor's inverse is then obtained as follows,
\begin{gather}
g^{\mu\nu}=\left( \begin{array}{cccc}
-\frac{1}{\Delta }f_{\phi\phi} & 0 & 0 & \frac{g_{c\phi}}{\Delta \sin^2 \theta} \\
0 & \frac{\Delta }{\rho^2} & 0 & 0 \\
0 & 0 & \frac{1}{\rho^2} & 0 \\
\frac{g_{c\phi}}{\Delta \sin^2 \theta} & 0 & 0 & {\frac{1}{\Delta {\sin }^{{\rm 2}}\theta} \left(1-\frac{r_s\left(r-d\right)-r^2_D}{\rho^2}\right)\ } \end{array}
\right),\\
f_{\phi\phi}=r\left(r-2d\right)-k^2+a^2+\frac{r_s\left(r-d\right)-r^2_D}{\rho^2}a^2{{\sin }^{{\rm 2}} \theta\ }.
\end{gather}

\section{The Klein-Gordon Equation, its solutions, and the quantization of the quasibound states}

In the present Section we formulate first the Klein-Gordon equation in the Dyonic Kerr-Sen black hole geometry, and then we obtain its exact solutions. The quantization of the quasibound states is also realized by using the properties of the Confluent Heun function.

\subsection{The Klein-Gordon equation}

We aim to investigate the massive and massless scalar fields in the curved spacetime generated by a Dyonic Kerr-Sen black hole, which is the solution of the  EMDA theory in the low energy limit of the heterotic string field theory. We will first construct, and then search for the exact solutions of the governing relativistic matter wave equation. We begin with writing the generic form of the Klein-Gordon in a curved spacetime background as follows,
\begin{equation}
-{\hbar }^2\nabla_\mu\nabla^\mu \psi+m^2 c^2 \psi=0,
\end{equation}
where $\nabla_\mu$ is the covariant derivative with respect to the metric.

For the covariant derivatives, we obtain,
$
\nabla_\mu\nabla^\mu \psi={\nabla }_\mu{\partial }^\mu\psi={\partial }_\mu{\partial }^\mu\psi+{\Gamma }^\mu_{\mu\nu}{\partial }^\nu\psi.
$
With the use of the identity ${\Gamma }^\alpha_{\alpha\beta}=\left(1/\sqrt{-g}\right)\left(\partial \sqrt{-g}/\partial x^\beta\right)$,
we can express the d'Alembert operator $\nabla_\mu\nabla^\mu \psi$ as $\nabla_\mu\nabla^\mu \psi=\left(1/\sqrt{-g}\right){\partial }_\mu\left(\sqrt{-g}g^{\mu\nu}{\partial }_\nu\psi\right)$.

Now, the Klein-Gordon equation can be expressed in terms of partial derivatives and the inverse metric tensor components as follows,
\begin{gather}
\left\{-{\hbar }^2\left[\frac{1}{\sqrt{-g}}\left({\partial }_\mu\sqrt{-g}g^{\mu \nu}{\partial }_\nu\right)\right]+m^2 c^2\right\}\psi=0,
\end{gather}
where $m$ is the rest mass of the scalar.

As the metric determinant and the metric inverse have already been obtained, the Laplace-Beltrami operator of the Klein-Gordon equation can be found component by component as follows,
\begin{widetext}
\begin{gather}
\frac{1}{\sqrt{-g}}{\partial }_0\sqrt{-g}g^{00}{\partial }_0{\rm =-}\frac{{\rm 1}}{\Delta \rho^2}\left\{{\left[r\left(r-2d\right)-k^2+a^2\right]}^2-\Delta a^2{{\sin }^{{\rm 2}} \theta\ }\right\}{\partial }^2_{ct},    \\
\frac{1}{\sqrt{-g}}{\partial }_0\sqrt{-g}g^{03}{\partial }_3{\rm =-}\frac{\left[r\left(r-2d\right)-k^2+a^2-\Delta \right]a}{\Delta \rho^2}{\partial }_{ct}{\partial }_\phi,  \\
\frac{1}{\sqrt{-g}}{\partial }_{{\rm 3}}\sqrt{-g}g^{30}{\partial }_0{\rm =-}\frac{\left[r\left(r-2d\right)-k^2+a^2-\Delta \right]a}{\Delta \rho^2}{\partial }_{ct}{\partial }_\phi,\\
\frac{1}{\sqrt{-g}}{\partial }_{{\rm 1}}\sqrt{-g}g^{11}{\partial }_1{\rm =}\frac{1}{\rho^2}{\partial }_r\left(\Delta {\partial }_r\right),
\end{gather}
\begin{gather}
\frac{1}{\sqrt{-g}}{\partial }_{{\rm 2}}\sqrt{-g}g^{22}{\partial }_2{\rm =}\frac{1}{\rho^2{\sin  \theta\ }}{\partial }_\theta\left({\sin  \theta\ }{\partial }_\theta\right),\\
\frac{1}{\sqrt{-g}}{\partial }_{{\rm 3}}\sqrt{-g}g^{33}{\partial }_3{\rm =}\frac{\Delta -a^2{{\sin }^{{\rm 2}} \theta\ }}{\Delta {{\sin }^2 \theta\ }\rho^2}{\partial }^2_\theta
\end{gather}
\end{widetext}

By combining all components, we obtain the explicit Klein-Gordon equation in the Dyonic Kerr-Sen black hole background as follows,
\begin{multline}
\left[{\rm -}\frac{{\rm 1}}{\Delta \rho^2}\left\{{\left[r\left(r-2d\right)-k^2+a^2\right]}^2-\Delta a^2{{\sin }^{{\rm 2}} \theta\ }\right\}{\partial }^2_{ct}\right. \\ \left.-2\frac{\left[r\left(r-2d\right)-k^2+a^2-\Delta \right]a}{\Delta \rho^2}{\partial }_{ct}{\partial }_\phi\right. \\ \left.+\frac{1}{\rho^2}{\partial }_r\left(\Delta {\partial }_r\right)+\frac{1}{\rho^2{\sin  \theta\ }}{\partial }_\theta\left({\sin  \theta\ }{\partial }_\theta\right)\right. \\ \left.+\frac{\Delta -a^2{{\sin }^{{\rm 2}} \theta\ }}{\Delta {{\sin }^2 \theta\ }\rho^2}{\partial }^2_\phi\right]\psi-\frac{m^2 c^2}{{\hbar }^2}\psi=0. \label{fullwave}
\end{multline}

The presence of temporal and azimuthal symmetry enables us to apply the separation ansatz \cite{35},
\begin{gather}
\psi\left(t,r,\theta,\phi\right)=e^{-i\frac{E}{c}ct+im_l\phi}R\left(r\right)T\left(\theta\right).
\end{gather}

\subsection{The Polar Wave Equation}

Let us define first the following dimensionless variables,
\begin{gather}
    \Omega=\frac{Er_s}{\hbar c} ,
    \Omega_0=\frac{E_0r_s}{\hbar c},
\end{gather}
where $E_0=mc^2$ is the scalar's rest energy.

Substituting the separation ansatz into the Eq.~\eqref{fullwave}, after multiplying the entire equation by $r^2/\psi \left(t,r,\theta ,\phi \right)$, we obtain,
\begin{multline}
\left\{{\rm -}\frac{{\rm 1}}{\Delta \rho^2}\left\{{\left(r\left(r-2d\right)-k^2+a^2\right)}^2-\Delta a^2{{\sin }^{{\rm 2}} \theta\ }\right\}\times \right. \\ \left.\left(-\frac{E^2}{\hbar^2 c^2}\right)-2\frac{\left(r\left(r-2d\right)-k^2+a^2-\Delta \right)a}{\Delta \rho^2}\left(\frac{E m_l}{\hbar c}\right)\right. \\ \left.+\frac{1}{R\rho^2}{\partial }_r\left(\Delta {\partial }_rR\right)+\frac{1}{T\rho^2{\sin  \theta\ }}{\partial }_\theta\left({\sin  \theta\ }{\partial }_\theta T\right)\right. \\ \left.+\frac{\Delta -a^2{{\sin }^{{\rm 2}} \theta\ }}{\Delta {{\sin }^2 \theta\ }\rho^2}\left(-m^2_l\right)\right\}-\frac{m^2 c^2}{{\hbar }^2}=0.
\end{multline}

By multiplying the entire wave equation by $\rho^2$, and after using the trigonometric identity ${{\sin }^{{\rm 2}} \theta\ }=1-{{\cos }^2 \theta\ }$, we obtain the following radial-polar equation,\\
\begin{multline}
\left[\frac{1}{T{\sin  \theta\ }}{\partial }_\theta\left({\sin  \theta\ }{\partial }_\theta T\right)-\frac{m^2_l}{{{\sin }^2 \theta\ }}-\left(\frac{\Omega^2_0 a^2}{r^2_s}-\frac{\Omega^2a^2}{r^2_s}\right){{\cos }^2 \theta\ }\right]\\+\left[\frac{1}{R}{\partial }_r\left(\Delta {\partial }_rR\right)+\frac{\Omega^2}{r^2_s}\frac{{\left(r\left(r-2d\right)-k^2+a^2\right)}^2}{\Delta }-\frac{\Omega^2a^2}{r^2_s}\right. \\ \left.-2\frac{\left(r\left(r-2d\right)-k^2+a^2-\Delta \right)a}{\Delta }\left(\frac{\Omega m_l}{r_s}\right)+\frac{m^2_l a^2}{\Delta }\right. \\ \left.-\frac{\Omega^2_0}{r^2_s}\left(r\left(r-2d\right)-k^2\right)\right]=0.
\end{multline}
The polar part can be separated as follows,
\begin{multline}
\frac{1}{T{\sin  \theta\ }}{\partial }_\theta\left({\sin  \theta\ }{\partial }_\theta T\right)-\frac{m^2_l}{{{\sin }^2 \theta\ }}\\-\left(\frac{\Omega^2_0 a^2}{r^2_s}-\frac{\Omega^2a^2}{r^2_s}\right){{\cos }^2 \theta\ }+\lambda^{m_l}_l=0.
\end{multline}

In the case of $a=k=0$, the separation constant is set to be $\lambda^{m_l}_l=l\left(l+1\right)$, and the polar wave solution is obtained in terms of the Legendre polynomial, $P^{m_l}_l (\cos \theta)$. But, in the case of a rotating black hole with a non-zero axion charge, the polar solution is obtained in terms of the Spheroidal function, $S^{m_l}_l\left(c_a,{\cos  \theta\ }\right)$, as follows \cite{NIST},
 \begin{equation}
\hspace{-0.4cm}T(\theta)=S^{m_l}_l\left(c_a,{\cos  \theta\ }\right)=\sum^{\infty }_{r=-\infty }{d^{\;lm_l}_r\left(c_a\right)P^{m_l}_{l+r}\left({\cos  \theta\ }\right)},
\end{equation}
where,
\begin{equation}
 c_a=   \frac{\Omega^2_0 a^2}{r^2_s}-\frac{\Omega^2a^2}{r^2_s}.
\end{equation}

\subsection{The Radial Wave Equation}

After solving the polar part, we are left with the following radial equation,
\begin{multline}
{\partial }_r\left(\Delta {\partial }_rR\right)+\left[\frac{\Omega^2}{r^2_s}\frac{{\left(r\left(r-2d\right)-k^2+a^2\right)}^2}{\Delta }-\frac{\Omega^2a^2}{r^2_s}\right. \\ \left.-2\frac{\left(r\left(r-2d\right)-k^2+a^2-\Delta \right)a}{\Delta }\left(\frac{\Omega m_l}{r_s}\right)+\frac{m^2_l a^2}{\Delta }\right. \\ \left.-\frac{\Omega^2_0}{r^2_s}\left(r\left(r-2d\right)-k^2\right)-\lambda^{m_l}_l\right]R{\rm =0} .   \label{radialeq}
\end{multline}

The radial equation needs a  careful treatment. First, remember that the condition ${\Delta}=0$ leads to two solutions, i.e., $r_\pm$. Hence, $\Delta$ can be rewritten in the factorized form as follows,
\begin{gather}
{\Delta}=(r-r_-)(r-r_+),
\end{gather}
giving
\begin{multline}
    \partial_r({\Delta}\partial_rR)=(r-r_-)\partial_rR\\+(r-r_+)\partial_rR+(r-r_-)(r-r_+)\partial^2_rR,
\end{multline}
and
\begin{gather}
\frac{r_+-r_-}{{\Delta}}=\frac{\delta_r}{{\Delta}}=\frac{1}{r-r_+}-\frac{1}{r-r_-},
\end{gather}
respectively. Rearranging Eq.~\eqref{radialeq}, we obtain,
\begin{multline}
{\partial }_r\left(\Delta {\partial }_rR\right)+\left[\frac{1}{\Delta }{\left\{\frac{\Omega}{r_s}\left(r\left(r-2d\right)-k^2+a^2\right)-m_la\right\}}^2\right. \\ \left.-\left\{\frac{\Omega^2_0}{r^2_s}\left(r\left(r-2d\right)-k^2+a^2\right)+\frac{\Omega^2}{r^2_s}a^2-\frac{\Omega^2_0}{r^2_s}a^2\right.\right. \\ \left.\left.-2\frac{\Omega m_l}{r_s}a+\lambda^{m_l}_l\right\}\right]R{\rm =0}, \label{radialmod}
\end{multline}
where the following constant, $K^{m_l}_l$, has been defined as,
\begin{gather}
K^{m_l}_l=\frac{\Omega^2}{r^2_s}a^2-\frac{\Omega^2_0}{r^2_s}a^2-2\frac{\Omega m_l}{r_s}a+\lambda^{m_l}_l.
\end{gather}

Working out the double radial differentiation in the first term of the equation \eqref{radialmod}, followed by multiplying the entire equation by $\delta_r^2$, we obtain the following equation,
\begin{widetext}
\begin{multline}
\delta^2_r{\partial }^{{\rm 2}}_rR+\left\{\frac{1}{r-r_+}+\frac{1}{r-r_-}\right\}\delta^2_r{\partial }_rR+\left[{\left(\frac{1}{r-r_+}-\frac{1}{r-r_-}\right)}^2{\left\{\frac{\Omega}{r_s}\left(r\left(r-2d\right)-k^2+a^2\right)-m_la\right\}}^2\right. \\ \left.-\delta_r\left(\frac{1}{r-r_+}-\frac{1}{r-r_-}\right)\left\{\frac{\Omega^2_0}{r^2_s}\left(r\left(r-2d\right)-k^2+a^2\right)+K^{m_l}_l\right\}\right]R=0.
\end{multline}
\end{widetext}

The region of interest of the investigation is outside the outer horizon, i.e. $r_+\leq r <\infty$. Hence, let us define the following new radial variable,
\begin{gather}
\delta_ry=r-r_+\to \delta_rdy=dr,\\
r-r_-=\delta_r\left(y+1\ \right).
\end{gather}

The domain is now shifted to $0 \leq y < \infty$. In terms of $y$, the radial equation becomes,
\begin{equation}
{\partial }^{{\rm 2}}_yR+\left\{\frac{1}{y}+\frac{1}{y+1}\right\}{\partial }_yR+\left[{{F.T.}^2} + S.T.\right]R{\rm =0}.
\end{equation}
where,
\begin{multline}
F.T. = \frac{1}{\delta_r}\left(\frac{1}{y}-\frac{1}{y+1}\right)\times\\
\left\{\frac{\Omega}{r_s}\left({\left(\delta_ry+r_+\right)}^2-2d\left(\delta_ry+r_+\right)-k^2+a^2\right)-m_la\right\},
\end{multline}
and,
\begin{multline}
S.T. = -\left(\frac{1}{y}-\frac{1}{y+1}\right)\times\\\left\{\frac{\Omega^2_0}{r^2_s}\left({\left(\delta_ry+r_+\right)}^2-2d\left(\delta_ry+r_+\right)-k^2+a^2\right)+K^{m_l}_l\right\}.
\end{multline}

Let us consider now the term $F.T.$, which can be rewritten as
\begin{multline}
F.T.
=\frac{1}{\delta_r}\frac{1}{y\left(y+1\right)}\left\{\frac{\Omega}{r_s}\left[\delta^2_ry^2+2y\delta_r(r_+ -d)\right]\right. \\ \left.+\underbrace{\frac{\Omega}{r_s}\left[r_+\left(r_+-2d\right)-k^2+a^2\right]-m_la}_{K_1}\right\}.
\end{multline}

We need to decompose the terms such as $\frac{y}{y+1}$ by using the fractional decomposition, $\frac{y}{y+1}=1-\frac{1}{y+1}$. Hence we continue as follows,
\begin{multline}
F.T.
=\frac{\Omega}{r_s}\delta_r+\frac{1}{y}\frac{K_1}{\delta_r}\\+\frac{1}{y+1}\left\{\underbrace{\frac{\Omega}{r_s}\left(r_++r_- -2d\right)-\frac{K_1}{\delta_r}}_{{K_3}}\right\}.
\end{multline}
After some algebraic transformations we obtain,
\begin{multline}
{F.T.^2}=\frac{\Omega^2}{r^2_s}\delta^2_r+\frac{K^2_1}{\delta^2_ry^2}+\frac{K^2_3}{{\left(y+1\right)}^2}+\frac{2\Omega K_1}{r_sy}\\
+\frac{2\Omega\delta_rK_3}{r_s\left(y+1\right)}+\frac{2K_1K_3}{\delta_r}\left(\frac{1}{y}-\frac{1}{y+1}\right).
\end{multline}

Now let us proceed to the calculation of $S.T.$,
\begin{multline}
S.T. =-\frac{1}{y\left(y+1\right)}\left\{\frac{\Omega^2_0}{r^2_s}\left(\delta^2_ry^2+2\delta_ry\left(r_+ -d\right)\right)\ \right. \\ \left.+\underbrace{\frac{\Omega^2_0}{r^2_s}\left(r_+\left(r_+-2d\right)-k^2+a^2\right)+K^{m_l}_l}_{K_2}\right\}.
\end{multline}
Again, applying the fractional decomposition, $\frac{y}{y+1}=1-\frac{1}{y+1}$, we find,
\begin{multline}
S.T.=-\frac{\Omega^2_0}{r^2_s}\delta^2_r-\frac{K_2}{y}\\+\frac{1}{y+1}\left[\underbrace{\frac{\Omega^2_0}{r^2_s}\delta^2_r-2\delta_r(r_+ -d)\frac{\Omega^2_0}{r^2_s}+K_2}_{K_4}\right].
\end{multline}

Up to this step, $F.T.^2$ and $S.T.$ have successfully been expressed in terms of $\frac{1}{y+1}$ and $\frac{1}{y}$. Let us rewrite the fully decomposed radial equation,
\begin{multline}
{\partial }^{{\rm 2}}_yR+\left[\frac{1}{y}+\frac{1}{y+1}\right]\partial_yR+\left[\left(\frac{\Omega^2}{r^2_s}\delta^2_r-\frac{\Omega^2_0}{r^2_s}\delta^2_r\right)\right. \\ \left.+\frac{1}{y}\left(\frac{2\Omega K_1}{r_s}+\frac{2K_1K_3}{\delta_r}-K_2\right)\right. \\ \left.+\frac{1}{y+1}\left(\frac{2\Omega \delta_rK_3}{r_s}-\frac{2K_1K_3}{\delta_r}+K_4\right)\right. \\ \left.+\frac{1}{y^2}\left(\frac{K^2_1}{\delta^2_r}\right)+\frac{1}{(y+1)^2}K^2_{3}\right]y=0.
\end{multline}

By redefining the radial variable $x=-y$, we obtain,
\begin{multline}
{\partial }^{{\rm 2}}_xR+\left[\frac{1}{x}+\frac{1}{x-1}\right]\partial_xR+\left[\left(\frac{\Omega^2}{r^2_s}\delta^2_r-\frac{\Omega^2_0}{r^2_s}\delta^2_r\right)\right. \\ \left.-\frac{1}{x}\left(\frac{2\Omega K_1}{r_s}+\frac{2K_1K_3}{\delta_r}-K_2\right)\right. \\ \left.-\frac{1}{x-1}\left(\frac{2\Omega \delta_rK_3}{r_s}-\frac{2K_1K_3}{\delta_r}+K_4\right)\right. \\ \left.+\frac{1}{x^2}\left(\frac{K^2_1}{\delta^2_r}\right)+\frac{1}{(x-1)^2}K^2_{3}\right]y=0. \label{radnat}
\end{multline}

The radial equation above is in the natural general form of the asymmetrical Confluent Heun equation \eqref{naturalheun}. Direct comparison between \eqref{radnat} and \eqref{naturalheun} allows us to identify $A_{1,2,3,4,5}$ as follows,\\
\begin{gather}
    A_1=-\left(\frac{2\Omega K_1}{r_s}+\frac{2K_1K_3}{\delta_r}-K_2\right),\\
    A_2=-\left(\frac{2\Omega \delta_rK_3}{r_s}-\frac{2K_1K_3}{\delta_r}+K_4\right),\\
    A_3=\frac{K^2_1}{\delta^2_r},\\
    A_4=K^2_3,\\
    A_5=\frac{\Omega^2}{r^2_s}\delta^2_r-\frac{\Omega^2_0}{r^2_s}\delta^2_r.
\end{gather}

Following the results of Appendix \ref{AppendixA}, the exact solutions of the radial equation \eqref{radnat} are obtained in terms of the Confluent Heun functions as follows,
\begin{multline}
y=e^{i\sqrt{A_5}}x^{i\sqrt{A_3}}(x-1)^{i\sqrt{A_4}}\left[A\operatorname{HeunC}\left(\alpha ,\beta ,\gamma ,\delta ,\eta ,x\right)\right. \nonumber\\ \left.+Bx^{-\beta}\operatorname{HeunC}\left(\alpha ,-\beta ,\gamma ,\delta ,\eta ,x\right)\right],
\end{multline}
where according to (A.16-A.18), we can express $i\sqrt{A_{3,4,5}}$ in terms of $\alpha,\beta,\gamma$ as follows,
\begin{gather}
\pm i\sqrt{A_3}=\frac{1}{2}\beta,\\
\pm i\sqrt{A_4}=\frac{1}{2}\gamma,\\
\pm i\sqrt{A_5}=\frac{1}{2}\alpha.
\end{gather}

The Confluent Heun function's parameters, following (A.15)-(A.19), are then explicitly obtained as follows,
\begin{gather}
\alpha =\pm 2i\frac{\delta_r}{r_s}\sqrt{\Omega^2-\Omega^2_0},\\
\beta=\pm \frac{{\rm 2}i}{\delta_r}\left[\frac{\Omega}{r_s}\left(r_+\left(r_+-2d\right)-k^2+a^2\right)-m_la\right], \label{beta}\\
{\gamma}=\pm \frac{2i}{\delta_r}\left[\frac{\Omega}{r_s}\left(r_-\left(r_--2d\right)-k^2+a^2\right)-m_la\right],\\
\delta=-\frac{\delta_r}{r^2_s}\left(r_++r_- -2d\right)\left[2\Omega^2-\Omega^2_0\right],
\end{gather}
\begin{multline}
\eta=-\frac{2}{\delta^2_r}\left(\frac{\Omega}{r_s}\left(r_+\left(r_+-2d\right)-k^2+a^2\right)-m_la\right)\times\\
\left(\frac{\Omega}{r_s}\left(-r_+^2+2r_-\left(r_+ -d\right)-k^2+a^2\right)-m_la\right)\\-\frac{\Omega^2_0}{r^2_s}\left(r_+\left(r_+-2d\right)-k^2+a^2\ \right)-K^{m_l}_l.
\end{multline}

Finally, we can present the complete exact solution of the scalar field's wave function in Dyonic Kerr-Sen black hole background as follows,
\begin{multline}
\psi =e^{-i{\frac{E}{\hbar c}}ct}S_{l}^{m_l}\left(\theta,\phi\right)e^{-\frac{1}{2}\alpha \left(\frac{r-r_+}{\delta_r}\right)}{\left(\frac{r-r_-}{\delta_r}\right)}^{\frac{1}{2}\gamma}\\
\times \left[A{\left(\frac{r-r_+}{\delta_r}\right)}^{\frac{1}{2}\beta}\operatorname{HeunC}\left(\alpha ,\beta ,\gamma ,\delta ,\eta ,-\frac{r-r_+}{\delta_r}\right)\right. \\ \left.+B{\left(\frac{r-r_+}{\delta_r}\right)}^{-\frac{1}{2}\beta}\operatorname{HeunC}\left(\alpha ,-\beta ,\gamma ,\delta ,\eta ,-\frac{r-r_+}{\delta_r}\right)\right],
\end{multline}\label{wavefunctions}
where,
\begin{equation}
 S_{l}^{m_l}\left(\theta,\phi\right)=e^{im_l \phi} S_{l}^{m_l}\left(\frac{\Omega^2_0 a^2}{r^2_s}-\frac{\Omega^2a^2}{r^2_s},\cos \theta\right).
\end{equation}

\subsection{QBS Energy Quantization}

The radial quantization condition is obtained from the degree of the interpolating radial function. The radial quantum number $n$ is defined as the number of zeros of the radial wave. The Confluent Heun function will have $n$ zeros if it is a polynomial function with degree $n$, and this condition is fulfilled when (see \ref{AppendixA}),
\begin{gather}
\frac{\delta}{\alpha}+\frac{\beta +\gamma}{2}=-n, \label{quanfor}
\end{gather}
where we have redefined $n=n_r+1=1,2,...$.

\subsubsection{The Quasistationary Modes} \label{energylevels}

In this Subsection, we will present the energy quantization expression \eqref{quanfor} for all possible quasistationary modes, which corresponds to combinations of positive and negative solutions of $\alpha,\beta,\gamma$. We will use the notation $X_{\pm}$, where $X$ could be $\alpha,\beta,\gamma$ and $\pm$ corresponds to the positive or negative solution of $X$.

1. For the case with $\alpha_+,\beta_+,\gamma_+$
the quantization condition \eqref{quanfor} can be rewritten explicitly as follows,
{\begin{multline}
-\frac{{\left(r_++r_- -2d\right)}\left[2\Omega^2-\Omega^2_0\right]}{2r_s\sqrt{\Omega^2_0-\Omega^2}}
+\frac{i}{\delta_r}\left\{\frac{\Omega}{r_s}\left(r_+(r_+-2d)\right. \right.\\ \left.\left.+r_-\left(r_--2d\right)+2\left(a^2-k^2\right)\right)-2m_la\right\}=-n.
\end{multline}}

2. For the case with $\alpha_+,\beta_+,\gamma_-$, we obtain,
{\begin{equation}
-\frac{{\left(r_++r_- -2d\right)}\left[-\Omega^2_0+2\Omega^2+2i\Omega\sqrt{\Omega^2_0-\Omega^2}       \right]}{2r_s\sqrt{\Omega^2_0-\Omega^2}}=-n.
\end{equation}}

3. For the case with $\alpha_+,\beta_-,\gamma_+$
, we obtain,
{\begin{equation}
-\frac{{\left(r_++r_- -2d\right)}\left[-\Omega^2_0+2\Omega^2-2i\Omega\sqrt{\Omega^2_0-\Omega^2}       \right]}{2r_s\sqrt{\Omega^2_0-\Omega^2}}=-n.
\end{equation}}

4. For the case with $\alpha_+,\beta_-,\gamma_-$, we obtain,
{\begin{multline}
-\frac{{\left(r_++r_- -2d\right)}\left[2\Omega^2-\Omega^2_0\right]}{2r_s\sqrt{\Omega^2_0-\Omega^2}}
-\frac{i}{\delta_r}\left\{\frac{\Omega}{r_s}\left(r_+(r_+-2d)\right. \right.\\ \left.\left.+r_-\left(r_--2d\right)+2\left(a^2-k^2\right)\right)-2m_la\right\}=-n.
\end{multline}}

5. For the case with $\alpha_-,\beta_+,\gamma_+$, we obtain,
{\begin{multline}
\frac{{\left(r_++r_- -2d\right)}\left[2\Omega^2-\Omega^2_0\right]}{2r_s\sqrt{\Omega^2_0-\Omega^2}}
+\frac{i}{\delta_r}\left\{\frac{\Omega}{r_s}\left(r_+(r_+-2d)\right. \right.\\ \left.\left.+r_-\left(r_--2d\right)+2\left(a^2-k^2\right)\right)-2m_la\right\}=-n.
\end{multline}}

6. For the case with $\alpha_-,\beta_+,\gamma_-$
, we obtain,
{\begin{equation}
\frac{{\left(r_++r_- -2d\right)}\left[-\Omega^2_0+2\Omega^2-2i\Omega\sqrt{\Omega^2_0-\Omega^2}       \right]}{2r_s\sqrt{\Omega^2_0-\Omega^2}}=-n.
\end{equation}}

7. For the case with $\alpha_-,\beta_-,\gamma_+$
, we obtain,
{\begin{equation}
\frac{{\left(r_++r_- -2d\right)}\left[-\Omega^2_0+2\Omega^2+2i\Omega\sqrt{\Omega^2_0-\Omega^2}       \right]}{2r_s\sqrt{\Omega^2_0-\Omega^2}}=-n.
\end{equation}}

8. For the case with $\alpha_-,\beta_-,\gamma_-$, we obtain,
{\begin{multline}
\frac{{\left(r_++r_- -2d\right)}\left[2\Omega^2-\Omega^2_0\right]}{2r_s\sqrt{\Omega^2_0-\Omega^2}}
-\frac{i}{\delta_r}\left\{\frac{\Omega}{r_s}\left(r_+(r_+-2d)\right. \right.\\ \left.\left.+r_-\left(r_--2d\right)+2\left(a^2-k^2\right)\right)-2m_la\right\}=-n.
\end{multline}}

Note that the scalar's energy levels in a rotating black hole spacetime possess hyperfine splitting, where the energy depends on the azimuthal quantum number $m_l$ and is directly coupled to the black hole's angular momentum parameter $a$. This is analogous to the Zeeman effect, which occurs when a hydrogen atom is immersed in a magnetic environment. The existence of the term can be understood as an interaction between the orbiting scalar field, with magnetic state $m_l$, and the black hole's angular momentum $a$.

Now, let us investigate the energy levels in the limit $(E-E_{0})r_s\to 0$. For modes with $\beta_+,\gamma_+$ and $\beta_-,\gamma_-$, we obtain,
\begin{gather}
E_n=E_0\sqrt{1-{\left[\frac{\frac{E_0r_s}{\hbar c}\frac{{\left(r_++r_- -2d\right)}}{r_s}}{2\left(i\frac{2m_la}{\delta_r}+n\right)}\right]}^2},\\
E_n-E_0\approx -\frac{E_0}{2}{\left[\frac{\frac{E_0r_s}{\hbar c}\frac{{\left(r_++r_- -2d\right)}}{r_s}}{2\left(i\frac{2m_la}{\delta_r}+n\right)}\right]}^2. \label{ens1}
\end{gather}
while for the cases with $\beta_-,\gamma_+$ and $\beta_+,\gamma_-$, we obtain,
\begin{gather}
E_n=E_0\sqrt{1-{\left[\frac{\frac{E_0r_s}{\hbar c}\frac{{\left(r_++r_- -2d\right)}}{r_s}}{2n}\right]}^2},\\
E_n-E_0\approx -\frac{E_0}{2}{\left[\frac{\frac{E_0r_s}{\hbar c}{\frac{{\left(r_++r_- -2d\right)}}{r_s}}}{2n}\right]}^2. \label{ens2}
\end{gather}

In contrast to the quasibound states around static black holes, where the energy levels are real valued in the small black hole limit \cite{Daniel,Anal1,Anal2,Anal3,Anal10,David-Senjaya}, for a rotating black hole, there are modes where complex valued energy levels are obtained even in a small black hole limit. The non zero decay of the quasibound state even in the small black hole limit in the rotating black hole backgrounds is expected, since, classically, an astrophysical rotating black hole with $a=0.998\frac{r_s}{2}$ has $32\%$ efficiency in converting the mass accreted from the disk to radiation. This value is five times greater than that of the non rotating Schwarzschild black hole \cite{Hobson}. Considering the non-rotating case, by taking as zero the black hole's angular momentum, $a=0$, we obtain the  expression of the energy levels  of the real valued quasi-bound states of the static axionic dyonic dilatonic Reissner-Nordstrom black hole.

Now, let us consider a massless scalar field in the Dyonic Kerr-Sen black hole's gravitational potential. Setting $\Omega_0=0$, we can derive the energy expressions for each modes as follows,

1. For the case with $\alpha_+,\beta_+,\gamma_+$, we obtain,
{\begin{gather}
\Omega_1=\frac{r_s\left(in(r_+-r_-)+2m_la\right)}{2\left(r_+(r_+-2d)+a^2-k^2\right)}, \label{mzero1} \\
\Omega_2=\frac{r_s\left(in(r_+-r_-)+2m_la\right)}{2\left(r_-(r_- -2d)+a^2-k^2\right)}.
\end{gather}}

2. For the case with $\alpha_+,\beta_+,\gamma_-$
, we obtain,
{\begin{equation}
\Omega=-\frac{inr_s}{2(r_++r_-)-4d}.
\end{equation}}

3. For the case with $\alpha_+,\beta_-,\gamma_+$
, we obtain,
{\begin{equation}
\Omega=\frac{inr_s}{2(r_++r_-)-4d}.
\end{equation}}

4. For the case with $\alpha_+,\beta_-,\gamma_-$, we obtain,
{\begin{gather}
\Omega_1=\frac{r_s\left(-in(r_+-r_-)+2m_la\right)}{2\left(r_+(r_+-2d)+a^2-k^2\right)},\\
\Omega_2=\frac{r_s\left(-in(r_+-r_-)+2m_la\right)}{2\left(r_-(r_- -2d)+a^2-k^2\right)}.
\end{gather}}

5. For the case with $\alpha_-,\beta_+,\gamma_+$, we obtain,
{\begin{gather}
\Omega_1=\frac{r_s\left(in(r_+-r_-)+2m_la\right)}{2\left(r_+(r_+-2d)+a^2-k^2\right)},\\
\Omega_2=\frac{r_s\left(in(r_+-r_-)+2m_la\right)}{2\left(r_-(r_- -2d)+a^2-k^2\right)}.
\end{gather}}

6. For the case with $\alpha_-,\beta_+,\gamma_-$
, we obtain,
{\begin{equation}
\Omega=-\frac{inr_s}{2(r_++r_-)-4d}.
\end{equation}}

7. For the case with $\alpha_-,\beta_-,\gamma_+$
, we obtain,
{\begin{equation}
\Omega=\frac{inr_s}{2(r_++r_-)-4d}.
\end{equation}}

8. For the case with $\alpha_-,\beta_-,\gamma_-$, we obtain,
{\begin{gather}
\Omega_1=\frac{r_s\left(-in(r_+-r_-)+2m_la\right)}{2\left(r_+(r_+-2d)+a^2-k^2\right)},\\
\Omega_2=\frac{r_s\left(-in(r_+-r_-)+2m_la\right)}{2\left(r_-(r_- -2d)+a^2-k^2\right)}.  \label{mzero8}
\end{gather}}

Notice that the cases with $\beta_-,\gamma_+$ and $\beta_+,\gamma_-$ have purely imaginary energy levels, indicating a more rapid absorption of the field  by the Dyonic Kerr-Sen black hole. There is also an apparent degeneracy in $\alpha$, since the solutions with $\alpha_+$ are the same as the solutions with $\alpha_-$.

\subsection{Numerical Investigations of the Black Hole's Parameters}

To investigate how the spin and charges of the Dyonic Kerr-Sen black hole affect a scalar field, let us choose an appropriate mode to investigate, i.e., the mode $\alpha_+,\beta_-,\gamma_-$, which has $\text{Re}(\Omega)>0$ that represents particle state.  We investigate the QNMs by plotting the exact energy expression for various spin and charges ranged from zero to near-extremal configurations~($r_{-}\sim r_{+}$). 

\begin{figure}[hbt!]
    \centering
    \includegraphics[scale=.9]{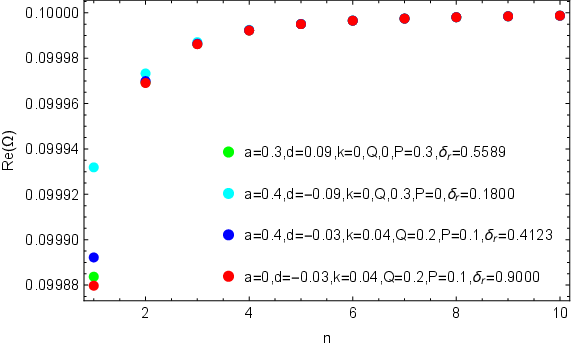}\\
    \includegraphics[scale=.9]{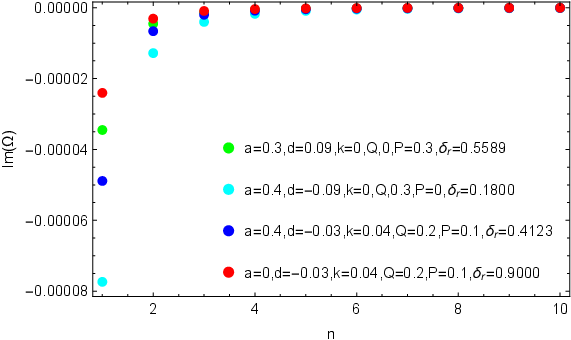}
    \caption{QNMs for various combinations of sub-extremal spin and charges.} \label{fig1}
\end{figure}

In Fig.~\ref{fig1}, we present the real and imaginary parts of the quasiresonance~(quasinormal) frequencies of $n=1,2,...,7$ states for various combinations of sub-extremal spin and charges (in geometrical unit), where the scalar and black hole mass are set to be respectively $r_s=1,\Omega_0=0.1$. Notice that the higher the excitation, the closer $\text{Re}(\Omega)$ to $\Omega_0$, indicating weaker binding energy for higher excited states.

\begin{figure}[hbt!]
    \centering
    \includegraphics[scale=.8]{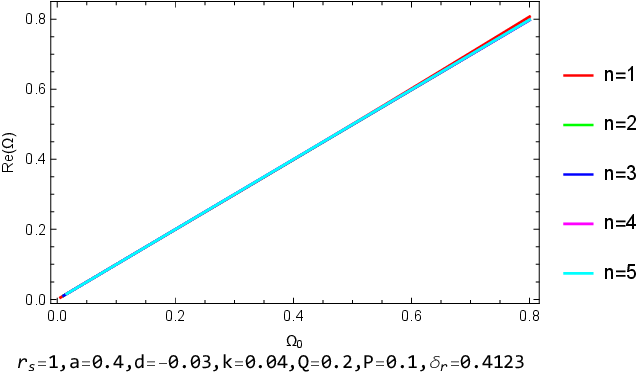}\\
    \includegraphics[scale=.8]{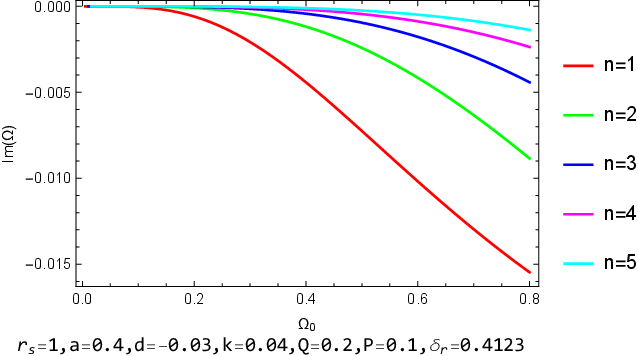}
    \caption{QNMs for varied scalar mass.}
  \label{fig2}
\end{figure}

\begin{figure}[hbt!]
    \centering
    \includegraphics[scale=.8]{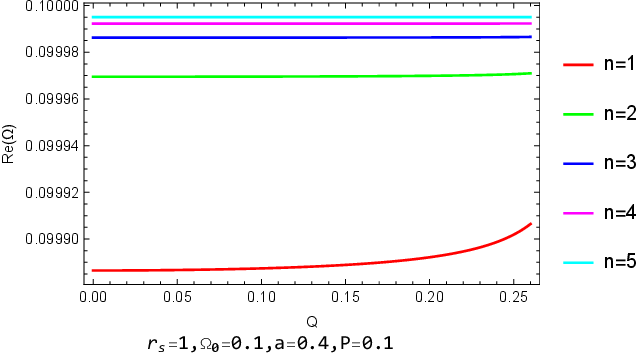}\\
    \includegraphics[scale=.8]{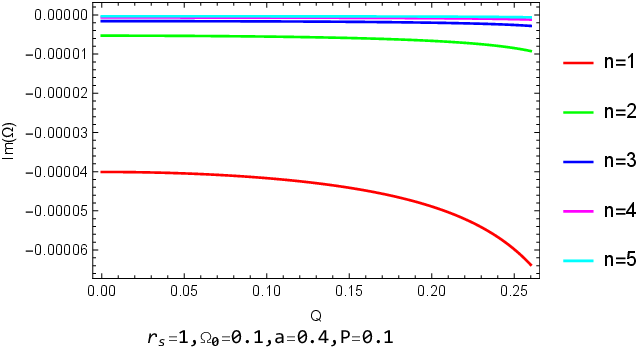}
    \caption{QNMs for varied electric charge.}
  \label{fig3}
\end{figure}

\begin{figure}[hbt!]
    \centering
    \includegraphics[scale=.8]{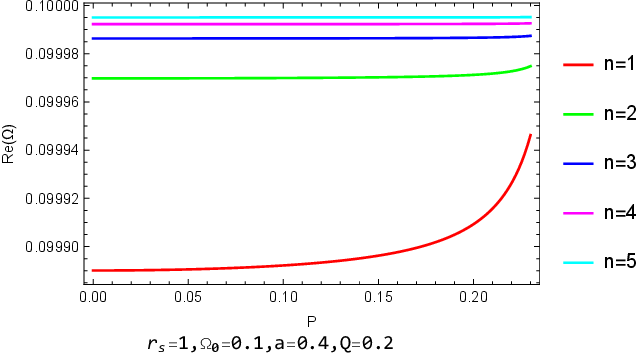}\\
    \includegraphics[scale=.8]{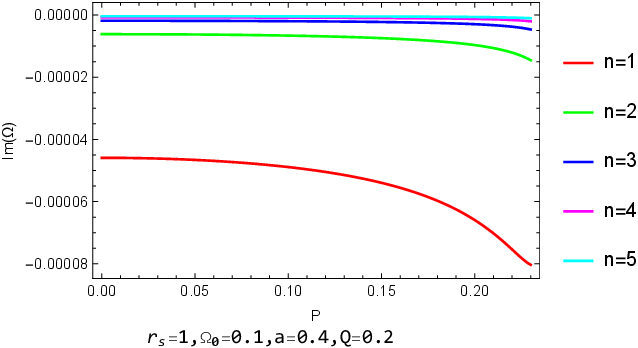}
    \caption{QNMs for varied magnetic charge.}
  \label{fig4}
\end{figure}

\begin{figure}[hbt!]
    \centering
    \includegraphics[scale=.8]{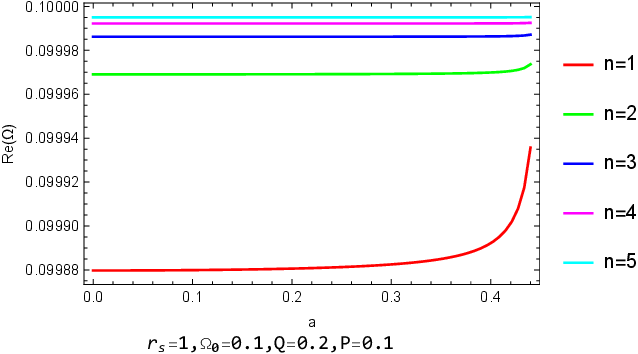}\\
    \includegraphics[scale=.8]{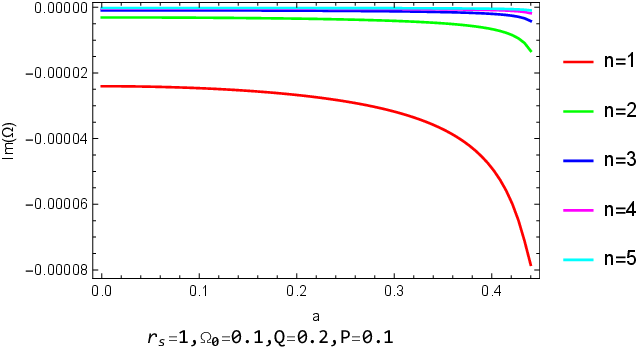}
    \caption{QNMs for varied angular momentum.}
  \label{fig5}
\end{figure}

In Fig.~\ref{fig2},\ref{fig3},\ref{fig4},\ref{fig5}, visualization of quasiresonance frequencies for $n=1,2,\dots ,5$ for varied scalar mass, electric charge, magnetic charge and angular momentum,  respectively, are presented. For each investigation, one black hole's parameter is varied from zero up to its near-extremal limit, while the others are kept constant. In all cases, we observe significant changes happen in the region where the Dyonic Kerr-Sen black hole becomes nearly extremal. Also notice that the states with smaller radial quantum number $n$, i.e., low excited states, are affected more than the states with large $n$, i.e., higher excited states.

\begin{figure}[hbt!]
    \centering
    \includegraphics[scale=.9]{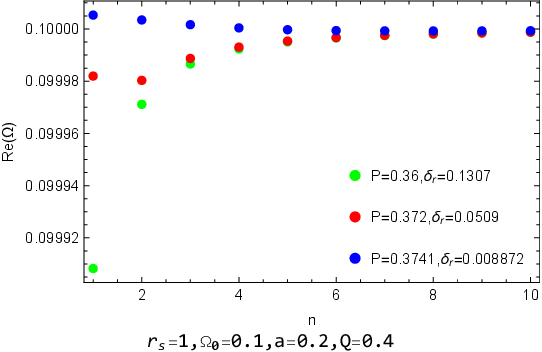}\\
    \includegraphics[scale=.9]{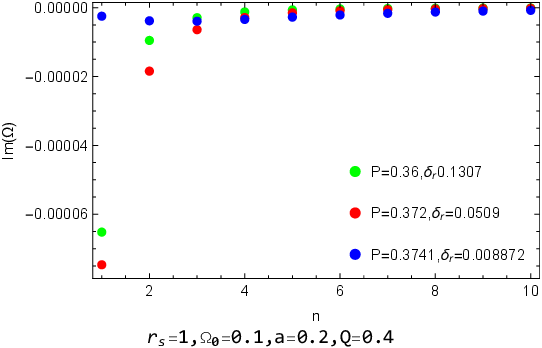}
    \caption{Near extremal QNMs for varied magnetic charge.}
  \label{fig6}
\end{figure}

\begin{figure}[hbt!]
    \centering
    \includegraphics[scale=.9]{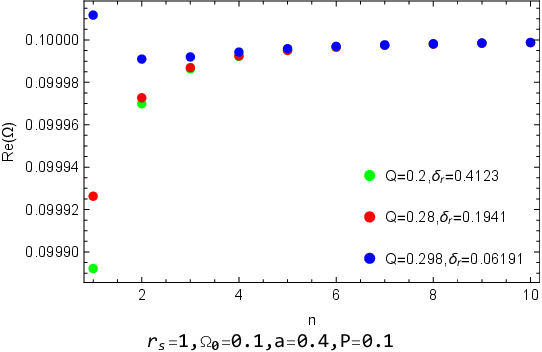}\\
    \includegraphics[scale=.9]{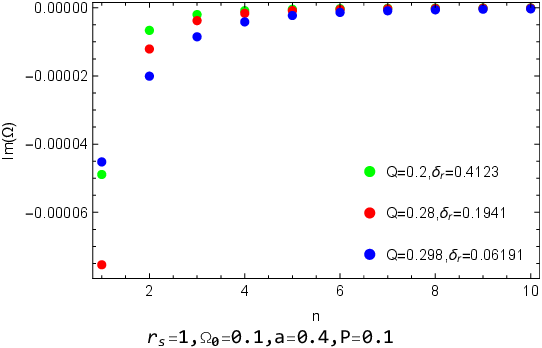}
    \caption{Near extremal QNMs for varied electric charge.}
  \label{fig7}
\end{figure}

\begin{figure}[hbt!]
    \centering
    \includegraphics[scale=.9]{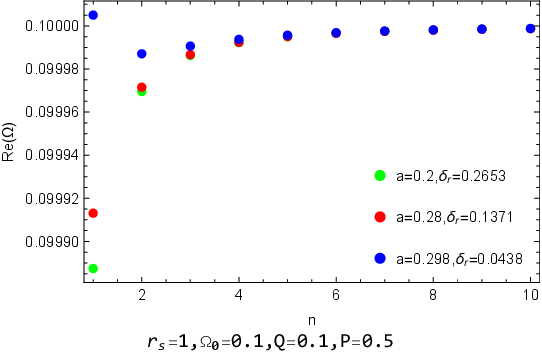}\\
    \includegraphics[scale=.9]{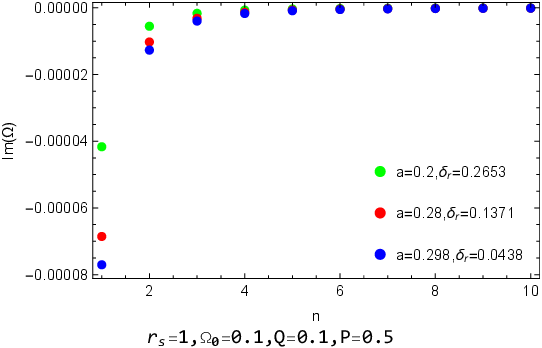}
    \caption{Near extremal QNMs for varied angular momentum.}
  \label{fig8}
\end{figure}

In Fig.~\ref{fig6},\ref{fig7},\ref{fig8}, we investigate how the QNM profile changes as the  Dyonic Kerr-Sen black hole becomes nearly extremal. Firstly, we fix the black hole's mass, spin and the electric charge and we plot the quasiresonance frequencies with respect to the magnetic charge approaching the extremal limit followed by varying the electric charge and angular momentum. We consistently observe significant changes in states with smaller radial quantum number $n$ as the black hole becomes nearly extremal. Also notice that very close to the extremal limit, we observe there are fundamental states having $\text{Re}(\Omega)$ larger than $\Omega_0$ (see Fig.~\ref{fig6},\ref{fig7},\ref{fig8}.)

\subsection{Analytical Investigations of the  Black Hole's Parameters}

In this subsection, we will investigate analytically, how the black hole's angular momentum and charges affect the scalar field's energy levels. The exact expression of the energy levels, in general, are complex valued quartic equations. It is very difficult to algebraically figure out how the black hole spin and charges affect the scalar's energy levels. However, it is possible to analytically investigate the approximated energy expression \eqref{ens1} in comparison with the well-known Schwarzschild's gravitational atom expression in \cite{Daniel, Anal1, Anal2, Anal3, Anal10, David-Senjaya, Huang} as follows,
\begin{equation}
{\left(E-E_0\right)}_{Sch}=-\frac{E_0}{8}{\left[\frac{E_0r_S}{\hslash c}\right]}^2{\left(\frac{1}{n}\right)}^2,
\end{equation}

Since the Schwarzschild black hole is the Dyonic Kerr-Sen black hole with vanishing charges and angular momentum, the comparison between the two may give some insights on how the black hole spin and charges affect the scalar field's quantized energy.

Now, let us consider the approximated energy expression in equation \eqref{ens1}. This particular mode is important and interesting due to its dependence on the black hole's angular momentum. The expression can be rewritten in $x+iy$ form as follows,\\
\begin{align}
{\left(E-E_0\right)}_{DKS}&\approx -\frac{E_0}{2}{\left[\frac{\frac{E_0r_S}{\hslash c}\left(\frac{r_++r_--2d}{r_s}\right)}{2\left(i\frac{2m_la}{{\delta }_r}+n\right)}\right]}^2\nonumber \\&\approx -\frac{E_0}{8}{\left[\frac{E_0r_S}{\hslash c}\left(\frac{r_++r_--2d}{r_s}\right)\right]}^2\times \nonumber \\&\phantom{=}\left(\frac{n^2-{n_a}^2}{{\left(n^2+{n_a}^2\right)}^2}-i\frac{2n_an}{{\left(n^2+{n_a}^2\right)}^2}\right),
\end{align}
where $n_a=\frac{2m_la}{{\delta }_r}$ and by using the relation \eqref{rpm}, we obtain $r_++r_--2d=r_s$.

Evaluating the ratio between ${\left(E-E_0\right)}_{Sch}$ and ${\left(E-E_0\right)}_{DKS}$, we obtain the following relation,
\begin{align}
\frac{{\left(E-E_0\right)}_{DKS}}{{\left(E-E_0\right)}_{Sch}}&=\ n^2\left(\frac{n^2-{n_a}^2}{{\left(n^2+{n_a}^2\right)}^2}-i\frac{2n_an}{{\left(n^2+{n_a}^2\right)}^2}\right)\nonumber\\&=\frac{1}{{\left(1+{\left(\frac{n_a}{n}\right)}^2\right)}^2}\left[1-{\left(\frac{n_a}{n}\right)}^2-2\frac{n_a}{n}i\right]\nonumber\\&={\mathfrak{R}}_{Re}+i{\mathfrak{R}}_{Im} ,
\end{align}
where,
\begin{equation}
{\mathfrak{R}}_{Re}=\frac{1-{\left(\frac{n_a}{n}\right)}^2}{{\left(1+{\left(\frac{n_a}{n}\right)}^2\right)}^2}\ \ \ ,\ \ \ {\mathfrak{R}}_{Im}=-2\frac{\frac{n_a}{n}}{{\left(1+{\left(\frac{n_a}{n}\right)}^2\right)}^2}.
\end{equation}

\begin{figure}[hbt!]
    \centering
    \includegraphics[scale=.9]{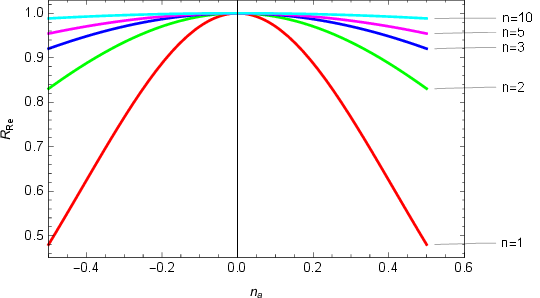}\\
    \includegraphics[scale=.9]{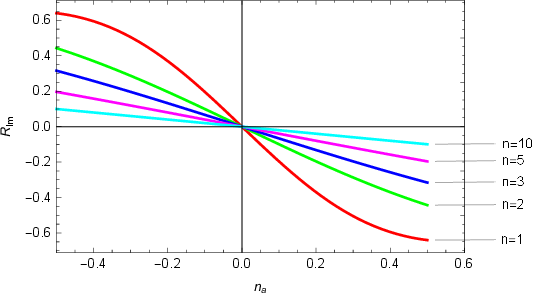}
    \caption{The ratio between scalar's binding energy in the Dionic Kerr-Sen with the Schwarzschild black hole in $(E-E_0)r_s \to 0$ limit.}
  \label{fig9}
\end{figure}

The function ${\mathfrak{R}}_{Re}$ and ${\mathfrak{R}}_{Im}$ for various value of excitation numbers are shown in Fig.~\ref{fig9}.

The variable ${\delta }_r=r_+-r_-$ which measures the distance between the outer and inner horizons has the following explicit expression,
\begin{equation}
{\delta }_r=2\sqrt{{\left(\frac{r_s}{2}\right)}^2+d^2+k^2-\left(a^2+r^2_D\right)}.
\end{equation}

Notice that the black hole's angular momentum together with electric and magnetic charges in $r^2_D$ are in opposite sign to the mass, dilaton and axion charges. This indicates that the presence of dilaton and axion charges shifts the inner horizon further inward and the outer horizon further outward resulted in larger separation between $r_+$ and $r_-$, while the presence of the spin affects the horizon separation in the opposite way. The parameter ${\delta }_r$ is then scaling the magnetic-spin interaction  $n_a=\frac{2m_la}{{\delta }_r}$, that larger ${\delta }_r$ reduces $n_a$ and vice versa, and the special case with $m_l=0$ or $a=0$ will effectively null the interaction term. 

The ratio $\displaystyle{\frac{{\left(E-E_0\right)}_{DKS}}{{\left(E-E_0\right)}_{Sch}}}$ tells us that for a scalar field in a general Dyonic Kerr-Sen space-time having main quantum number $n$ and non-zero magnetic quantum number $m_l$, the real part of ${\left(E-E_0\right)}_{DKS}$ will always be smaller than the same mode in the Schwarzschild space-time. There will also be additional decay for modes with $m_l>0$ and instability for modes with $m_l<0$. Thus, the presence of the black hole's angular momentum and all charges affect fundamental mode, i.e. $n=1$, more than modes with larger $n$ since,
\begin{equation}
{\mathop{\mathrm{lim}}_{n\to \infty } {\mathfrak{R}}_{Re}\ }=1,\ \ \ \ \ {\mathop{\mathrm{lim}}_{n\to \infty } {\mathfrak{R}}_{Im}\ }=0.
\end{equation}

For the massless scalar case given by (\ref{mzero1})-(\ref{mzero8}), the presence of non-zero black hole’s spin gives rise to QBS with non-zero real part, i.e. solutions for underdamped modes with $\beta_+,\gamma_+$ and $\beta_-,\gamma_-$. while the rest are not affected by the black hole’s spin and charges and only depend only on the black hole’s mass since $r_++r_--2d=r_s$.

\section{The Horizon's Hawking Radiation}

In the previous Section, we have presented the detailed derivations for obtaining the complete expressions of the polar and radial waves in terms of the Spheriodal Harmonics and of the Confluent Heun functions. In this Section, we focus on the Hawking radiation from the apparent horizon of the Dyonic Kerr-Sen black hole. Following the Sigurd - Sannan method \cite{Sannan:1988eh}, we can use the exact radial wave solutions to derive the Hawking radiation distribution function, and search for the expression of the Hawking temperature of the black hole's apparent horizon $r_+$. Approaching the exterior event horizon $r\to r_+$, we can approximate $\operatorname{HeunC}$ as follows,
\begin{equation}
\operatorname{HeunC}(0)=\operatorname{HeunC}'(0)\approx1,
\end{equation}
also the exponential, $e^{-\frac{1}{2}\alpha \left(\frac{r-r_+}{\delta_r}\right)}\approx 1$,
\begin{gather}
R ={\left(\frac{r_+-r_-}{\delta_r}\right)}^{\frac{1}{2}\gamma}\left[B{\left(\frac{r-r_+}{\delta_r}\right)}^{-\frac{1}{2}\beta}+A{\left(\frac{r-r_+}{\delta_r}\right)}^{\frac{1}{2}\beta}\right],\\
\beta=\frac{{\rm 2}i}{\delta_r}\left[\frac{\Omega}{r_s}\left(r_+\left(r_+-2d\right)-k^2+a^2\right)-{m_la}\right].
\end{gather}

The radial wave consists of two independent parts,
\begin{gather}
R =\left\{ \begin{array}{cc}
\psi_{+in}=A{\left(\frac{r_+-r_-}{\delta_r}\right)}^{\frac{1}{2}\gamma}{\left(\frac{r-r_+}{\delta_r}\right)}^{-\frac{1}{2}\beta} & \emph{ingoing} \\
\psi_{+out}=B{\left(\frac{r_+-r_-}{\delta_r}\right)}^{\frac{1}{2}\gamma}{\left(\frac{r-r_+}{\delta_r}\right)}^{\frac{1}{2}\beta} & \emph{outgoing} \end{array}
\right..
\end{gather}

Suppose there is an ingoing wave hitting the apparent horizon $r_+$. This will induce a particle-antiparticle pair, with the particle being reflected, while the antiparticle will be transmitted, going through the horizon. The analytical continuation of the wave function can be calculated as follows,
\begin{equation}
\begin{split}
{\left(\frac{r-r_+}{\delta_r}\right)}^{\lambda }&={\left(\frac{r_+}{\delta_r}\right)}^{\lambda }{\left(\frac{r}{r_+}-1\right)}^{\lambda }\\&\to {{\left(\frac{r_+}{\delta_r}\right)}^{\lambda }\left[\left(\frac{r}{r_+}-1\right)-i\epsilon \right]}^{\lambda }\\
&=\left\{ \begin{array}{cc}
{\left(\frac{r_+}{\delta_r}\right)}^{\lambda }{\left(\frac{r-r_+}{\delta_r}\right)}^{\lambda } & ,\ r>r_+ \\
{\left(\frac{r_+}{\delta_r}\right)}^{\lambda }{\left\lvert \frac{r-r_+}{\delta_r}\right\rvert }^{\lambda }e^{-i\lambda \pi} & ,\ r<r_+ \end{array}
\right.,
\end{split}
\end{equation}

The analytical continuation enables us to obtain the expression $\psi_{-out}=\psi_{+out}\left(\left(\frac{r-r_+}{\delta_r}\right)\to \left(\frac{r-r_+}{\delta_r}\right)e^{-i\pi}\right)$ simply by  $\left(\frac{r-r_+}{\delta_r}\right)\to -\left(\frac{r-r_+}{\delta_r}\right)=\left(\frac{r-r_+}{\delta_r}\right)e^{-i\pi}$ as follows,
\begin{equation}
\begin{split}
\psi_{-out}&=B{\left(\frac{r_+-r_-}{\delta_r}\right)}^{\frac{1}{2}\gamma}{\left(\left(\frac{r-r_+}{\delta_r}\right)e^{-i\pi}\right)}^{\frac{1}{2}\beta},\\
&=\psi_{+out}e^{-\frac{1}{2}i\pi \beta}. \label{anac}
\end{split}
\end{equation}

One can also find the modulus square of the probability amplitude with respect to the ingoing wave as follows,
\begin{align}
{\left\lvert \frac{\psi_{-out}}{\psi_{+in}}\right\rvert }^2&={\left\lvert \frac{\psi_{+out}}{\psi_{+in}}\right\rvert }^2e^{-i2\pi \beta}\\
&={\left\lvert \frac{\psi_{+out}}{\psi_{+in}}\right\rvert }^2e^{\frac{4\pi}{\delta_r}\left[\frac{E}{\hslash c}\left(r_+\left(r_+-2d\right)-k^2+a^2\right)-{m_la}\right]}\\&={\left\lvert \frac{\psi_{+out}}{\psi_{+in}}\right\rvert }^2 e^\zeta.  \label{analiticalcon}
\end{align}

The exponent $e^{-\zeta}$ represents the relative probability of radiation emission. The amplitude of the pair productions that occur is described by that function. Since the observer stays outside the black hole horizon, the absolute probability of the processes occurring outside the horizon needs to be found by summing all probabilities to create no pair, 1 pair, 2 pairs and so on, as follows,
\begin{equation}
  C_\omega\left(1+e^{-\zeta}+{\left(e^{-\zeta}\right)}^2+\dots \right)=1\to C_\omega=1-e^{-\zeta}.
\end{equation}

The probability to create $j$ pairs of particle-antiparticle is given by,
\begin{equation}
C_\omega {\left(e^{-\zeta}\right)}^j=\left(1-e^{-\zeta}\right)e^{-j\zeta}.
\end{equation}

The normalized spectrum of all the possible pair productions is obtained by calculating the mean number of the emitted particles,
\begin{equation}
n\left(\omega\right)=\sum^{\infty}_{n=0}n\left(1-e^{-\zeta}\right)e^{-n\zeta}=\frac{1}{e^{\zeta}-1}.  \label{distfunc}
\end{equation}

The same distribution function can also be obtained via the normalization condition of the wave function, following the method developed by \cite{Zhao}. Now, let us write the total outgoing wave that consists of the particle wave outside the black hole's horizon and the antiparticle wave inside the black hole's horizon. With the help of the Heaviside step function, the total outgoing wave can be written in a unique form as follows,
\begin{align}
\psi_{out}=\psi_{+out}\Theta \left(r-r_+\right)+\psi_{-out}\Theta \left(r_+-r\right),
\end{align}
or,
\begin{align}
\frac{\psi_{out}}{\psi_{+in}}=\frac{\psi_{+out}}{\psi_{+in}}\Theta \left(r-r_+\right)+\frac{\psi_{-out}}{\psi_{+in}}\Theta \left(r_+-r\right),
\end{align}

The total probability of the emission of the particles and of the antiparticles (remember that the antiparticle has a negative probability) must be normalized to be one. Hence, we can write,
\begin{equation}
\left\langle{\left\lvert \frac{\psi_{out}}{\psi_{+in}}\right\rvert }^2\right\rangle = \left\langle{\left\lvert \frac{\psi_{+out}}{\psi_{+in}}\right\rvert }^2\right\rangle-\left\langle{\left\lvert \frac{\psi_{-out}}{\psi_{+in}}\right\rvert }^2\right\rangle=1,
\end{equation}
and by using \eqref{analiticalcon} to substitute ${\left\lvert \frac{\psi_{-out}}{\psi_{+in}}\right\rvert }^2$, we obtain the following expression,
\begin{gather}
\left\langle{\left\lvert \frac{\psi_{+out}}{\psi_{+in}}\right\rvert }^2\right\rangle\left\lvert 1-e^{\zeta}\right\rvert=1, \\
\left\langle{\left\lvert \frac{\psi_{+out}}{\psi_{+in}}\right\rvert }^2\right\rangle=\frac{1}{e^{\zeta}-1}.
\end{gather}

The Hawking temperature, $T_H$, is to be found by first rewriting $\zeta$ as follows,
\begin{align}
\zeta&=\frac{4\pi}{\delta_r}\left[\frac{E}{\hslash c}\left(r_+\left(r_+-2d\right)-k^2+a^2\right)-{m_la}\right]\nonumber\\&
=\frac{4\pi}{\delta_r}\left[\frac{\omega }{c}\left(r_+\left(r_+-2d\right)-k^2+a^2\right)-{m_la}\right]\nonumber\\
&=\frac{\hslash \left(\omega -{\omega }_J\right)}{\left[\frac{\delta_rc\hslash }{4\pi \left(r_+\left(r_+-2d\right)-k^2+a^2\right)}\right]},
\end{align}
where we have defined,
\begin{equation}
{\omega }_J=\frac{m_l a c}{r_+\left(r_+-2d\right)-k^2+a^2}.
\end{equation}

Comparing the black hole's radiation distribution function \eqref{distfunc} with the Bose-Einstein distribution function,
 \begin{equation}
n\left(\omega\right)=\frac{1}{e^{\frac{\hbar\omega-\mu}{k_B T}}-1},
\end{equation}
the apparent horizon's Hawking temperature is found as follows,
\begin{gather}
T_H=\frac{\delta_rc\hslash }{4\pi k_B\left[r_+\left(r_+-2d\right)-k^2+a^2\right]}.\label{hawking}
\end{gather}

Interestingly enough, similar to the case of the Kerr black hole, the angular momentum plays the role of the chemical potential in the thermodynamics of Kerr-Sen black hole, with $\mu = \hbar\omega_{J}$  \cite{Navarro}. For the Kerr-Sen black hole, $\mu$ also depends on the axion and dilaton~(or electric and magnetic) charges of the black hole.

One can check that by setting the black hole's angular momentum and charges to be zero, we simultaneously set $r_+ \to r_s$, $r_- \to 0$ and $\delta_r \to r_s$. In this case, we recover the Hawking temperature of the Schwarzschild black hole \cite{Harris},
\begin{equation}
 T_H=\frac{c\hslash }{4\pi k_Br_s}.
\end{equation}

\section{Conclusions}

In this work, we have presented the novel exact analytical general solutions of the covariant massive and massless Klein-Gordon equations in the
Dyonic Kerr-Sen black hole spacetime.
By exactly solving the Klein-Gordon equation, we have also obtained the solutions for the canonical relativistic quantum mechanical problem of the energy levels of the scalar fields, gravitationally bounded by the Dyonic Kerr-Sen black hole. We have showed in detail the derivations of our results, and we have presented the  energy expressions of the all possible  sixteen quantize modes for both the massive and massless cases \ref{energylevels}, together with the corresponding exact wave functions \eqref{wavefunctions}. Since we did not use any approximations to derive the analytical solutions, the obtained wave functions are valid for all regions of interest, i.e. $r_+ \le r < \infty$. This is a remarkable improvement as compared to the previous approaches,  and, in contrast with the asymptotical methods, whose solutions are correct only for regions closed to the horizon, or asymptotically far away from the horizon.

We can recover the Schwarzschild massive quasibound state's real valued energy levels expression in the small black hole limit, given by,
\begin{gather}
\frac{E_n}{E_0}\approx 1-\frac{\kappa^2}{8n^2},   \\
\kappa={\left(\frac{E_0 r_s}{\hbar c}\right)}^2, \label{gravatom}
\end{gather}
by applying the small black hole limit $Er_s\to 0$ and taking as zero the spin $a$, and the charges of the Dyonic Kerr-Sen black holes, $Q,P,k,d$, in the equations \eqref{ens1} and \eqref{ens2}. The $\frac{1}{n^2}$ Hydrogenic-atom-like energy expression was also obtained in the previous works \cite{Daniel, Anal1, Anal2, Anal3, Anal10, David-Senjaya, Huang}.

We have performed further numerical and analytical investigations on how the Dyonic Kerr-Sen black hole's parameters affect the energy levels of the scalar fields. We plot graphical visualizations of QBS with various  configurations of scalar field-Dyonic Kerr-Sen black hole and find out that in general, the states with larger radial quantum number $n$ have $\text{Re}(\Omega)$ closer to $\Omega_0$, indicating weaker binding energy \ref{fig1}. Very close to the extremal limit, we observe there are states with small radial quantum number having $\text{Re}(\Omega)$ larger than $\Omega_0$ \ref{fig6},\ref{fig7},\ref{fig8}. In general, we find that states with smaller radial quantum number $n$ are more sensitive to the change of the black hole's parameters, which in agreement with the analytical investigation that has been done by making use of the approximated energy expression \eqref{ens1}.

In the last section, by making use of the exact radial solutions, we have investigated the Hawking radiation of the Dyonic Kerr-Sen black hole's outer horizon $r_+$. We have followed the Sigurd-Sannan method \cite{Sannan:1988eh,Damour} to treat the Klein pair production scenario as describing an incoming particle hitting  the black hole's horizon. The radiation distribution function is derived by summing all of the possible pair productions rates, and is presented in \eqref{distfunc}. Comparing it with the bosonic distribution function, the Hawking temperature of the Dyonic Kerr-Sen black hole's apparent horizon was obtained in \eqref{hawking}.

Analytical solutions, and results,  are extremely useful in the study of the astrophysical and physical properties of the black holes, including the study of the dynamics and motion of the  particles gravitating around them. They can also be successfully used to investigate the emissivity properties in various frequency ranges of the thin accretion disks that generally are present around black holes. The obtained exact solution of the scalar quasibound states of the Dyonic Kerr-Sen black hole can also help in discriminating this type of black hole with respect to standard general relativistic black holes, or other black hole solutions in modified gravity theories, and for obtaining observational constraints on the black hole parameters, and on mass, spin and the various charges of the black hole. 

We have also presented  a detailed investigation of the thermodynamic properties of the Dyonic Kerr-Sen black hole, by using the obtained exact analytical results.  The Hawking temperature is one of the extremely important physical parameters of the black holes, representing an essential quantity that has very important theoretical implications. In the limiting case of the vanishing angular momentum and charges we recover the expression of the Hawking temperature of the general relativistic Schwarzschild black holes. As compared to the standard Schwarzschild or Kerr cases, the horizon temperature of the Dyonic Kerr-Sen black holes has a strong dependence on the electric and magnetic charges, a new physical feature that could in principle discriminate between the various types of black holes, and their physical properties. The study of the thermodynamical properties of the black hole was very much facilitated by the existence of the exact solution. 

Bound systems play an important role in classical mechanics, being fundamental for the understanding  of the dynamics of the Sun and of the Solar System planets.  The study of the bound states of the quantum mechanical systems, like, for example, the hydrogen atom, opened new perspectives on the behavior of elementary particles. The discovery of the gravitational waves, together with the confirmation  of the existence of black holes by using various astrophysical methods and physical methods, led to a significant increase in the interest for bound and scattering states, and for quasinormal excitations of the particles gravitating around compact massive objects. There are no real bound states for particle around black holes, since they will cross into the black hole via quantum effects, leading to the slow, or rapid decay of the matter waves. Hence, these decaying states are called  quasibound states. Generally, for astrophysical black holes, the mass of the quasibound particles is extremely small,  and thus it is very difficult to detect them by using accelerator experiments. 

However, the situation is drastically different in an astrophysical environment, and the physical processes around black holes may lead to the possibility of testing the existence and properties of ultra light particles. If they have proper frequencies, the particles gravitating around a hole could create a large number of similar ones. The type of radiation corresponding to these processes (superradiance)  is a non-thermal one, and it is different from the Hawking radiation. Due to the black hole properties, some particles will cross the event horizon into the black hole by quantum tunnelling. However, if the two processes of creation and absorption are in detailed balance, a cloud of particles  can form  around the black hole \cite{Hod, Herdeiro}. The high spatial resolution, polarimetric imaging of supermassive black holes, like, for example,  M87* or SgrA* by the Event Horizon Telescope can be used to prove the existence of ultralight bosonic particles \cite{Chen}. The particles, existing around a rotating black hole due to the superradiance mechanism, concentrate in an accretion type disk. When linearly polarized photons are emitted from an accretion disk near the horizon, when traveling through the background cloud, their position angles oscillate due to the birefringent effect.  The periodic change of the position angle can be tested observationally both spatially and temporally. The detection of such oscillations could give strong evidence for the existence of superradiance \cite{Chen}. In this context the existence of the exact analytical solutions for the quasibound states could be very helpful for the understanding of the formation of the bosonic cloud, and of its properties. 

The ultra light particles gravitating around black holes also have an effect on the gravitational waves emitted by the black holes in binary systems. Interesting enough, even a single gravitational - wave measurement can detect the ultra light bosons located around the gravitational wave source \cite{Hann}. The observations of gravitational waves also show the existence of signals from the bosonic clouds formed by bound state of ultra light particles due to the spin-induced multipole moments and tidal Love numbers \cite{Bau}. 

Dyonic Kerr-Sen Black Hole have more variability as associated to their basic properties, as compared to the standard Schwarzschild or Kerr black holes, leading to a more complicated external dynamics. These richer properties do follow from the presence of the rotation, spin, and electric and magnetic charges, which also imply that the resulting black hole solution satisfies very complicated and strongly nonlinear field equations. As we have already seen, the effects associated with the spin and the charge degrees of freedom can lead to specific astrophysical effects and signatures, whose observational detection could open some new perspectives on the important relation between quantum and gravitational effects. In the present work, we have provided, via an exact solution of the Klein-Gordon equation in the Dyonic Kerr-Sen black hole background,
some of the basic tools necessary for a detailed comparison of the predictions of the string inspired Einstein-Maxwell-dilaton-axion theory of gravity with the results of astrophysical observations. These results could also lead to a deeper understanding of the nature of the string theoretical effects, and of their physical relevance.

\section*{Acknowledgments}
We would like to thank Teephatai Bunyaratavej for pointing out an error in the earlier version of the paper. DS acknowledges this work is supported by the Second Century Fund (C2F), Chulalongkorn University, Thailand. P.B. is supported in part by National Research Council of Thailand~(NRCT) and Chulalongkorn University under Grant N42A660500. This research has received funding support from the NSRF via the Program Management Unit for Human Resources \& Institutional Development, Research and Innovation~[grant number B41G670027].

\appendix

\section{The Confluent Heun Functions and S-Homotopic Transformation}
The Confluent Heun equation is a second-order linear differential equation that has three regular singularities given in the following canonical form \cite{Heun},  \label{AppendixA}
\begin{multline}
    \frac{d^2y_H}{dx^2}+\left(\alpha +\frac{\beta +1}{x}+\frac{\gamma +1}{x-1}\right)\frac{dy_H}{dx}
\\+\left(\frac{\mu }{x}+\frac{\nu }{x-1}\right)y_H=0, \label{canonincalheun}
\end{multline}
where
\begin{gather}
    \mu=\frac{1}{2}\left(\alpha-\beta-\gamma+\alpha\beta-\beta\gamma\right)-\eta,\\
    \nu =\frac{1}{2}\left(\alpha +\beta +\gamma +\alpha \gamma +\beta \gamma \right)+\delta +\eta.
 \end{gather}

The solutions of the differential equation are given in terms of two independent Confluent Heun functions as follows,
\begin{multline}
    y_H=A\operatorname{HeunC}\left(\alpha ,\beta ,\gamma ,\delta ,\eta ,x\right)\\+Bx^{-\beta}\operatorname{HeunC}\left(\alpha ,-\beta ,\gamma ,\delta ,\eta ,x\right).
\end{multline}

The Confluent Heun function can be reduced to an $n^{th}$ order polynomial function if the following series termination condition is fulfilled,
\begin{equation}
\frac{\delta}{\alpha}+\frac{\beta +\gamma}{2}+1=-n_r,\quad n_r\in\mathbb{Z}. \label{HeunPol}
\end{equation}

Given a natural general form of the asymmetrical Confluent Heun equation \cite{Heun,Vier22},
\begin{multline}
\frac{d^2y}{dx^2}+\left(\frac{1}{x}+\frac{1}{x-1}\right)\frac{dy}{dx}
\\+\left(\frac{A_1}{x}+\frac{A_2}{x-1}+\frac{A_3}{x^2}+\frac{A_4}{(x-1)^2}+A_5\right)y=0,  \label{naturalheun}
\end{multline}
in order to find the solution of \eqref{naturalheun}, we have to apply the s-homotopic transformation \cite{Alb,Heun,Vier22,Sylv,chen} by transforming the dependent variable $y(x)\to u(x)$ as follows,
\begin{equation}
y(x)=e^{B_0 x}x^{B_1}(x-1)^{B_2}u(x). \label{transf}
\end{equation}

Substituting the transformation into the equation \eqref{naturalheun}, we find the values of the exponents $B_0,B_1,B_2$ from the initial equation as follows,
\begin{gather}
B_0(B_0-1)+B_0+A_5=0 \to B_0=\pm i\sqrt{A_5},\\
B_1(B_1-1)+B_1+A_3=0 \to B_1=\pm i\sqrt{A_3},\\
B_2(B_2-1)+B_2+A_4=0 \to B_2=\pm i\sqrt{A_4}.
\end{gather}

Thus, after obtaining the three exponents, substitution of the s-homotopic transformation \eqref{transf} into \eqref{naturalheun} leads to a differential equation for $u(x)$ as follows,
\begin{multline}
    \frac{d^2u}{dx^2}+\left(2B_0+\frac{2B_1+1}{x}+\frac{2B_2+1}{x-1}\right)\frac{du}{dx}
\\+\left(\frac{\sigma}{x}+\frac{\chi}{x-1}\right)u=0,    \label{result}
\end{multline}
where
\begin{gather}
    \sigma=-B_1-B_2-2B_1B_2+B_0+2B_0B_1+A_1,\\
    \chi =B_1+B_2+2B_1B_2+B_0+2B_0B_2+A_2.
 \end{gather}

By comparing \eqref{result} with \eqref{canonincalheun}, we can write the solution for $u(x)$ in terms of the Confluent Heun functions as follows,
\begin{multline}
    u=A\operatorname{HeunC}\left(\alpha ,\beta ,\gamma ,\delta ,\eta ,x\right)\\+Bx^{-\beta}\operatorname{HeunC}\left(\alpha ,-\beta ,\gamma ,\delta ,\eta ,x\right),
\end{multline}
where,
\begin{gather}
    \alpha=2B_0=\pm 2i\sqrt{A_5},\\
    \beta=2B_1=\pm 2i\sqrt{A_3},\\
    \gamma=2B_2=\pm 2i\sqrt{A_4},\\
    \delta=A_1+A_2,\\
    \eta=-A_1.
\end{gather}

Hence, the complete solutions for the natural general form of the asymmetrical Confluent Heun equation \eqref{naturalheun} are obtained as follows,
\begin{multline}
y=e^{\pm i\sqrt{A_5}x}x^{\pm i\sqrt{A_3}}(x-1)^{\pm i\sqrt{A_4}}\left[A\operatorname{HeunC}\left(\alpha ,\beta ,\gamma ,\delta ,\eta ,x\right)\right. \nonumber\\ \left.+Bx^{-\beta}\operatorname{HeunC}\left(\alpha ,-\beta ,\gamma ,\delta ,\eta ,x\right)\right], \label{finalsol}
\end{multline}
with $\alpha,\beta,\gamma,\delta,\eta$ are given by (A.15)-(A.19).

For $x\to\infty$, the approximate solution of \eqref{canonincalheun} is given as follows,
\begin{align}
    y_H&=Ax^{-\left(\frac{\delta}{\alpha}+\frac{\beta+\gamma+2}{2}\right)}+Be^{-\alpha x}x^{\left(\frac{\delta}{\alpha}-\frac{\beta+\gamma+2}{2}\right)}\\
    &=e^{-\frac{1}{2}\alpha x}x^{-\frac{\beta+\gamma+2}{2}}\left[Ae^{\frac{1}{2}\alpha x}x^{-\frac{\delta}{\alpha}}+Be^{-\frac{1}{2}\alpha x}x^{\frac{\delta}{\alpha}}\right]\\
    &=y_0 e^{-\frac{1}{2}\alpha x}x^{-\frac{\beta+\gamma+2}{2}} \sin\left[-i\frac{\alpha}{2}x+i\frac{\delta}{\alpha}\ln{x}+\phi_0\right]\\
    &=y_0 e^{-\frac{1}{2}\alpha x}x^{-\frac{\delta}{\alpha}}x^{n_r} \sin\left[-i\frac{\alpha}{2}x+i\frac{\delta}{\alpha}\ln{x}+\phi_0\right].
\end{align}

\end{document}